# Two-stage optimization of urban rail transit formation and real-time station control at comprehensive transportation hub

Hualing REN[1*], Yingjie SONG[1], Shubin LI[2]

[1] School of Traffic and Transportation, Beijing Jiaotong University, China. * Corresponding author, hlren@bjtu.edu.cn.

[2] Department of Traffic Management Engineering, Shandong Police College, China.

**Abstract**: Urban rail transit connecting with comprehensive transportation hub should meet the passenger demand not only within the urban area, but also from outer area through high-speed railway or plane, which leads to the different characteristics of passenger demand. This paper tries to discuss two strategies of dealing with this complex passenger demand from two aspects: transit train formation and real-time holding control. First of all, we establish a model to optimize the multi-marshalling problem by minimizing the trains' vacant capacities when it leaves the station so as to cope with the fluctuation of demand in different periods. Then, we establish another model to control the multi-marshalling operated rail trains in real-time to relieve the passengers' delay caused by the fluctuated coming of other transportation modes from outer area, in which the trains' holding time at station is optimized aiming at minimizing the passengers' total waiting time. The genetic algorithm (GA) is designed to solve the integrated two-stage model of optimizing the number, timetable and real-time holding control of the multi-marshalling operated trains. The numerical results show that the combined two-stage model of multi-marshalling operation and holding control at stations can better deal with the demand fluctuation of urban rail transit connecting with the comprehensive transportation hub. This method can efficiently reduce the number of passengers detained at the hub station as well as the waiting time without increasing the passengers' on-train time even under the state of highly fluctuated passenger flow.

**Keywords**: multi-marshalling optimization, real-time holding control, comprehensive transportation hub, urban rail transit, genetic algorithm

## 1. Introduction

Comprehensive transportation hub connecting with multiple transportation modes is the key to the connecting and transferring of traffic between inside and outside the metropolis. The urban rail transit, which has been built and operated in many cities due to its advantages of large capacity, high efficiency, and punctuality, its stable operation connecting to the hub not only directly promotes the operation of various traffic modes, but also benefits the effective integration of urban internal and external traffic.

For urban rail transit line connected with hubs, passenger flow demand mainly comes from the daily commuter passenger flow within the city and outer transportation modes connected with the comprehensive transportation hubs. The daily commuter passenger flow inside the city has obvious tide phenomenon and fluctuates obviously in the whole day period; the arrival passenger flow of outer transportation modes connected with the hub presents intermittent short-term high-intensity characteristics. In addition, outer transportation modes connected with the hub are often delay due to weather or other reasons, resulting in randomness of the passenger arrival.

The current strategy of the rail transit company is adjusting the train departure time or headway during different time intervals of a day to match the transportation capacity supply with the fluctuated daily commuter passenger demand, that is, small interval for peak demand while large interval for off-peak demand. This makes much different waiting time for passengers arriving at the station at different time, especially during the period between the peak hour and the off-peak hour (Guo et al, 2017). As the existence of arrival passenger demand from outer transportation, this strategy is hard to deal with the complex passenger demand and balance the waiting time of different passengers throughout the day. In addition, when outer transportation modes arrive at random fluctuations because of delay, it is likely to cause an increasing of waiting time for some passengers.

The fluctuation of passenger demand at different periods affects a lot the operation efficiency of public transportation. In recent years, many scholars have done a lot of researches on public vehicle operation control and multi-vehicle operation organization to reduce the effect of the fluctuation of passenger demand and improve the service level of public transportation.

(1) Vehicle operation control

Vehicle operation control is an effective means to improve the level of public transportation service, especially for lines with high departure frequency, which makes real-time adjustments to the vehicle by continuously monitoring the operation status of the system and uses limited resources to best meet the needs of passenger travel. Among them, the most common method is the vehicle control at stations (Xu et al, 2001).

Early scholars did not consider the acquisition of real-time information of the system when modelling and analyzing the vehicle station control but presented the optimization model by controlling the specific station or specific vehicle. Newell and G. F (2016) minimized the average waiting time of passengers by controlling a pair of vehicles on a line. Barnett (1974) minimized passengers' total time of waiting at stations and on-board by controlling vehicles at specific stations.

Later, with the development of science and technology, the real-time state information of the system is taken in consideration in the process of vehicle operation control. Yu and Yang (2009) proposed a two-stage operation control strategy. In the first stage, according to the dynamic running time between stations, support vector machine was used to predict the vehicle departure time at the current and the next station, so as to determine the vehicle control strategy. In the second stage, the passengers' arrival rate is assumed to be constant, the vehicle control time is optimized to minimize the sum of passengers' waiting time at station and on-board time. According to the collected real-time location information of a vehicle, the influence of current station control on the length of vehicle delay and the departure intervals, Xu et al (2001) presented a deterministic quadratic programming model to minimize passengers' total waiting time. Zhao et al (2005) proposed a distributed control scheme, the vehicle's departure time from the station can be dynamically adjusted through the real-time communication between the vehicle and the station, aiming to minimize the sum of the waiting time at the station and the in-vehicle time.

Adamski and Turnau (1998) developed a series of vehicle control strategies based on the control theory, aiming to ensure that the vehicle run with the schedule; at the same time, they also proposed a variant of the control strategy that can balance the departure interval of the vehicle. Daganzo (2009) discussed the control of a pair of vehicles: stopping the following vehicle at the station when the headway is reduced and accelerating the following vehicle when the headway is increased. Bartholdi and Eisenstein (2012) adjusted the station control duration of the vehicle based on the departure interval between the current vehicle and the following vehicle to achieve balanced departure intervals.

Some studies combined vehicle station control with other control strategies. on the premise that the real-time information of the system can be obtained when each vehicle arrives at the station, Delgado et al (2009) combined vehicle station control and passenger number on board control under the constraint of vehicle capacity, and established a quadratic programming model aiming at minimal total travel time of passengers. Based on this research, Delgado et al (2012) compared the results of various control strategies under different combinations of arrival rate and running time between stations using simulation to verify the application conditions and control effect of the combined control strategy. Su and Wilson (2001) set up a mixed integer programming model to determine whether station control and regional vehicle operation should be carried out, assuming that passenger arrival rate and running time between stations are constant when the vehicle operation is slightly disturbed. Chandrasekar et al (2002) suggested implementing a signal priority to the front vehicle and a station control for the rear vehicle when the headway on the line is relatively small.

Some studies compared the effects of various control strategies by means of simulation. Grube and Cipriano (2010) proposed two real-time stop control strategies for subway lines to minimize passengers' waiting time. The first strategy only considered the vehicle and passenger flow information of the current station; the other one considered the vehicle and passenger flow information of the station within the prediction range based on the prediction model . Sanchez et al (2016) considered dynamic passenger arrival rate and running time between stations based on the system current status information, predicted the future state information, and presented a minimizing program to optimize vehicle station control time with the objective of minimizing waiting time during the period . Bellei and Gkoumas (2009) also analyzed the control effects of two-vehicle station control strategies on a bus line considering both dynamic passenger arrival rate and running time between stations,

(2) Multi-vehicle organization optimization

The research on multi-vehicle organization problem mainly considers the optimization of vehicle types and departure frequency under different demand scenarios and constraints, so as to meet the passenger flow demand in peak demand periods without causing waste of transport capacity in off-peak demand periods.

Some studies have optimized the size of buses based on the characteristics of demand. Jansson (1980) firstly considered issuing the same number of buses in different demand periods throughout the day, and then determined appropriate vehicle capacity according to the passenger flow in different periods under different requirements. Based on the relationship among vehicle size, operating costs and demand levels, Oldfield (1980) established a model to optimize vehicle size by maximizing social benefits and used the data from a British bus line for case analysis. Fu and

Ishkhanov (2004) presented a model of determining the optimal vehicle size under different conditions with different service levels, and analyzed the effects of multi-vehicle organization schemes. Tisato (2000) discussed the bus subsidy standards under different vehicle sizes from the perspective of economic analysis, and studied the relationship between vehicle capacity and cost. From practical point, Hassold and Ceder (2012) studied the selection of different vehicles with different number of seats considering passengers' distribution characteristics so that passengers' waiting time and empty mileage of vehicles are minimized.

Other studies combined bus size optimization with timetable optimization. Sun et al (2015) established a schedule optimization model based on multi-size bus model, and compared the passenger travel costs and company operating costs between multi-size bus model and single bus model. Olio (2012) established a two-layer programming model considering the constraint of the vehicle number: first layer assigns different sizes of buses to different routes to optimize the sum of passenger's travel costs and operator costs; second layer optimizes the departure frequency of each route according to the observed demand level. Yu et al. (2018) proposed a double-objective optimization model for the bus schedule and vehicle type selection considering the spatial-temporal distribution characteristics of passenger flow demand. Ceder and Dano (2013) presented a model to equalize both departure interval and load ratio of vehicles on a bus route: find optimal departure interval for given goal of load rate equalization with multiple vehicle types and optimize departure intervals of different types of vehicles. Kim and Schonfeld (2013) gave a model to optimize sizes of buses, scopes of the bus services, departure intervals and number of vehicles considering the combination of fixed buses and flexible buses.

In this paper, we focus on the optimal strategy to satisfy the passenger demand not only within the urban area, but also from outer transportation with different characteristics at the comprehensive transportation hub. As shown in Fig. 1, all these two kinks of coming passengers at the rail station can board on the $k^{th}$ train if there is no delay of the outer transportation; but they have to take the $(k+1)^{th}$ train because they arrive at the platform after the $k^{th}$ train leaves the station, and moreover, they may have to wait for the $(k+2)^{th}$ train because the $(k+1)^{th}$ train is over congested.

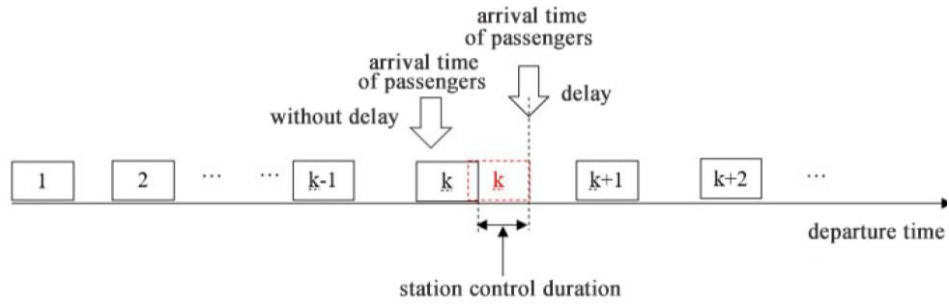

Fig 1. Operation control of rail train

To satisfy the passenger demands from both daily commuter and outer transportation, this paper proposes a two-stage strategy: adjust the train formation to deal with the short-term high-intensity demand instead of adjusting the train departure interval so as to remain a relative even headway and a relative equal average waiting time for all coming passengers at any time; control the holding times of the multi- marshalling trains to deal with the random fluctuation of both passenger demands. Therefore, the holding time of trains in this paper considers not only extending the time duration of passengers getting on and off the train, but also leaving ahead of time; that is, to optimize the dwelling time of the train at the station to take away at most passengers when the passengers' arrival time is within the preset time range. As shown in Fig. 1 above, a reasonable dwelling time should be given to the $k^{th}$ train when the passengers' delay is within the assumed time range so that the delayed passengers can catch up the $k^{th}$ train without any extra waiting time.

The main contributions of this paper include the following:

(1) It proposes a two-stage model to formulate the organization problem of rail transit connecting with comprehensive transportation hub when the passenger demands are from both inner urban area and outer transportation;

(2) The station control of train includes not only increasing the dwell time but also decreasing the dwell time, which means the train can also leave the station ahead of the schedule.

(3) The model considers not only the scheduled fluctuation of passenger demand caused by the scheduled arrival of outer transportation, but also the random fluctuation of passenger demand caused by fluctuated delay of the outer transportation to the hub due to different factors.

The remainder of this paper is organized as follows. In Section 2, we formulate the two-stage model of rail train organization. Then, the GA solving method is proposed inn Section 3 and Numerical examples are discussed in Section 4 to illustrate the properties of the proposed model and the performance of the algorithm. Finally, a summary and conclusions are presented in Section 5.

## 2. Model formulation

Assume that the urban rail transit line considered by the model has $N$ stations in total, and $K$ trains are sent out during the study period. Each train departs from station 1 to station $N$ and stops at each station. The symbols used in this article are listed below:

| | |
|---|---|
| $[t_s, t_e]$ | the time period; |
| $k$ | vehicle number ($k \in 1,2,...,K$); |
| $i,j$ | station number, $i,j \in 1,2,...,N$; |
| $v$ | train type, $v \in V$; |
| $x_k^v$ | 0-1 variable: if the $k^{th}$ train is train type $v$, it is 1; otherwise 0; |
| $c^v$ | passenger capacity of marshalling train $v \in V$; |
| $t_{k,i}^d$ | departure time of the $k^{th}$ train from station $i$; |
| $t_{k,i}$ | departure interval between the $k^{th}$ and $k+1^{th}$ train at station $i$; |
| $t_i^r$ | running time of the train between station $i$ and $i+1$; |
| $t_{k,1}^d$ | departure time of the $k^{th}$ vehicle at the first stop; |
| $t_{k,i}^{dw}$ | scheduled dwelling time of the $k^{th}$ train at station $i$; |
| $t^{min}, t^{max}$ | minimal and maximal bound of departure intervals for train operation; |
| $h^{max}$ | upper bound of time the train is allowed to ahead and behind the schedule; |
| $h_{k,i}$ | control time of the $k^{th}$ train at station $i$. |
| $\lambda_{i,j}(t)$ | dynamic passenger arrival rate from station $i$ to destination $j$; |
| $\xi(x,t)$ | random error function used to describe the random arrival delay of outer transportation at the hub; |
| $\alpha_{k,i}$ | number of passengers getting off the $k^{th}$ train at station $i$; |
| $\beta_{k,i,j}$ | passenger demand destined for destination $j$ at station $i$ when the $k^{th}$ train leaves station $i$; |
| $\overline{\beta_{k,i,j}}$ | number of passengers boarding on the $k^{th}$ train at station $i$ with the destination $j$; |
| $\beta_{k,i}$ | passenger demand when the $k^{th}$ train leaves station $i$; |
| $\overline{\beta_{k,i}}$ | actual number of boarding passengers when the $k^{th}$ vehicle leaves station $i$; |
| $\gamma_{k,i}$ | number of passengers on the $k^{th}$ vehicle when it leaves station $i$; |
| $l_{k,i}$ | number of passengers who arrive before the $k^{th}$ train leaves station $i$ and are left behind because of the lack of capacity on the train; |
| $c_{k,i}$ | spare capacity of the $k^{th}$ train when it leaves station $i$; |

**2.1 Model assumptions**

This paper studies one running direction of urban rail transit lines connecting with comprehensive transportation hub. Historical origin destination (OD) passenger flow along urban rail transit lines can be obtained using the auto fare collection (AFC) data[, and arrival schedules of outer transportation at the hub can also be obtained from transportation hub. In addition, the establishment of the model also requires the following three assumptions:

a). The running time between any two successive stations is fixed and determined by distance;

b). Trains of different marshalling groups have the same running properties (speed, acceleration, etc.), and each carriage has the same capacity in this paper.

c). Considering the train capacity (carriage capacity * number of groups), passengers follow the "first come first serve" principle when waiting at the station, and passengers wait for no more than two trains.

**2.2 Optimization objective**

The optimization model in this paper is completed in two stages. In the first stage, according to passenger arrival rate from outer transportation to the transportation hub and the time-varying OD demand between stations along the urban rail transit lines, the model adjusts the trains' departure time and the corresponding train formations under the constraint that the passengers will not wait for more than two trains. The object is to minimize the spare capacity on the train when the trains leave the stations during the period, and obtain the optimal multi-marshalling operating scheme and optimal running schedule of the rail transit trains. In the second stage, station control of the running trains is optimized to minimize the total waiting time of the passengers on the line during the period as the trains run according to the optimal schedule calculated in the first stage, considering the fluctuant coming of the outer transportation to the hub due to different factors.

Two objective functions of model are as follows:

(1) During the study period, the sum of spare capacity when all trains leave all stations is calculated as follows:

$$Z_1 = \sum_{k=1}^{K} \sum_{i=1}^{N} c_{k,i}, \tag{1}$$

where $c_{k,i}$ represents the spare capacity of the $k^{th}$ train when it leaves station $i$.

(2) Suppose that passengers wait for no more than two trains, so their waiting time can be divided into two parts: the necessary waiting time for the first train and possible waiting time for the second one. During the study period, the total waiting time of passengers at all stations on the line is calculated as follows:

$$Z_2 = T_1 + T_2. \tag{2}$$

where $T_1$ is the total waiting time for the first train, which is equal to the difference between the time of passengers arriving and the time of their first waiting train leaving the station:

$$T_1 = \sum_{k=1}^{K} \sum_{i=1}^{N} \sum_{j=i+1}^{N} \int_{t_{k-1,i}^d}^{t_{k,i}^d} \lambda_{i,j}(t)(t_{k,i}^d - t)dt. \tag{3}$$

where $\lambda_{i,j}(t)$ is dynamic passenger arrival rate from station $i$ to destination $j$, and we set $t_{0,i}^d = t_s$. $T_2$ is extra waiting time of some passengers for the second train due to the capacity constraint of their first waiting train. The value is the difference between the departure time of the next train and the current train:

$$T_2 = \sum_{k=2}^{K} \sum_{i=1}^{N} \sum_{j=i+1}^{N} l_{k,i,j}(t_{k,i}^d - t_{k-1,i}^d). \tag{4}$$

**2.3 Constraint conditions**

(1) Constraints in the first stage include

In the first stage, the constraints include

$$\sum_{v=1}^{V} x_k^v = 1, \ \forall k \in K, \tag{5}$$

$$t_s \leq t_{1,1}^d \leq t_s + t^{max}, \tag{6}$$

$$t_e - t^{max} \leq t_{K,1}^d \leq t_e, \tag{7}$$

$$t_{k,i} = t_{k+1,i}^d - t_{k,i}^d, \ \forall k \in K, i \in N, \tag{8}$$

$$t^{min} \leq t_{k,i} \leq t^{max}, \ \forall k \in K, i \in N, \tag{9}$$

$$t_{k,i+1}^d = t_{k,i}^d + t_i^r + t_{k,i}^{dw}, \ \forall k \in K, i \in N, \tag{10}$$

$$\beta_{k,i,j} = \int_{t_{k-1,i}^d}^{t_{k,i}^d} \lambda_{i,j}(t)dt, \quad \forall k \in K, i,j \in N, \tag{11}$$

$$\beta_{k,i} = \sum_{j=i+1}^{N} \beta_{k,i,j} + l_{k-1,i}, \quad \forall k \in K, i \in N, \tag{12}$$

$$\overline{\beta_{k,i}} = \min\left(\beta_{k,i}, \sum_{v=1}^{V} x_k^v \cdot c^v - \gamma_{k,i-1} + \alpha_{k,i}\right), \quad \forall k \in K, i \in N, \tag{13}$$

$$l_{k,i} = \beta_{k,i} - \overline{\beta_{k,i}}, \quad \forall k \in K, i \in N, \tag{14}$$

$$\overline{\beta_{k,J,i}} = \beta_{k,j,i} \cdot \frac{\overline{\beta_{k,J}}}{\beta_{k,j}}, \quad \forall k \in K, i,j \in N, \tag{15}$$

$$\alpha_{k,i} = \sum_{j=1}^{i-1} \overline{\beta_{k,J,i}}, \quad \forall k \in K, i,j \in N, \tag{16}$$

$$\gamma_{k,i} = \gamma_{k,i-1} + \overline{\beta_{k,i}} - \alpha_{k,i}, \quad \forall k \in K, i \in N, \tag{17}$$

$$c_{k,i} = \sum_{v=1}^{V} x_k^v \cdot c^v - \gamma_{k,i}, \quad \forall k \in K, i \in N, \tag{18}$$

$$\sum_{k=1}^{K-1} \sum_{i=1}^{N} (\overline{\beta_{k+1,i}} - l_{k,i}) \geq 0, \tag{19}$$

Eq. (5) indicates that each train corresponds to only one of the train types; Eqs. (6) and (7) limit the departure times of the first and last train in the study period; Eq. (9) shows that any departure time interval should be within the minimal and maximal departure interval; Eq. (11) refers to the number of passengers arriving at station $j$ from station $i$ when the $k^{th}$ vehicle leaves station $i$; Eq. (12) represents the number of waiting passengers when the $k^{th}$ train leaves station $i$. Eq. (13) shows the actual number of passengers on the $k^{th}$ train when it leaves station $i$; Eq. (14) is the number of passengers left behind due to the insufficient capacity.

The model assumes that the arriving passengers follow the first-come-first-serve principle, so when the $k^{th}$ train leaves station $i$, passengers from station $j$ with different destinations have the same probability of getting on the $k^{th}$ train, which is $\overline{\beta_{k,J}}/\beta_{k,j}$. Eq. (15) represents the number of passengers from station $j$ who can take the $k^{th}$ train when it leaves station $i$; Eq. (16) shows the number of passengers getting off the $k^{th}$ train when it arrives at station $i$; Eq. (17) is the number of passengers on board the $k^{th}$ train when it leaves station $i$; Eq. (18) is the spare capacity of the $k^{th}$ train when it leaves station $i$; Eq. (19) restricts passengers to wait for at most two trains.

(2) Constraints in the second stage

In the second stage of the model the station control of trains is optimized based on the multi-marshalling train operation schedule obtained in the first stage. The constraints in this stage include

Eqs. (7) - (11)

Eqs. (13) - (21)

$$t_{k,i+1}^d = t_{k,i}^d + t_i^r + t_{k,i}^{dw} + h_{k,i}, \quad \forall k \in K, i \in N, \tag{20}$$

$$-h^{max} \leq h_{k,i} \leq h^{max}, \quad \forall k \in K, i \in N. \tag{21}$$

Eq. (20) is the departure time of the train after station control and Eq. (21) is the feasible duration of station control.

## 3. Solution algorithm

Two genetic algorithms (GA) are adopted combinedly to solve the two-stage model in this paper. The algorithm steps are as follows:

**Step 0**: parameter initialization: the number of iterations gen = 0; the initial population size *M*; the algorithm termination algebra *N*.

**Step 1**: Perform chromosome coding and repeat it *M* times to get the initial population.

The gene sequence of the chromosome in Step 1 consists of three parts, as shown in Fig. 2. K can be repeatedly and randomly selected from *V* integers representing the group type as the first part of the chromosome; *K* real numbers that satisfy the constraints (6)-(9) are randomly generated as the second part of the chromosome; The $K*(N-2)$ randomly generated real numbers satisfying constrains (8)-(10) are taken as the third part of the chromosome.

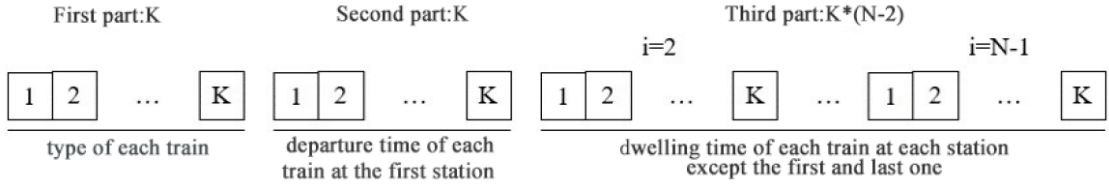

Fig 2. Chromosome coding for step 1

**Step 2**: Calculate the individual fitness value: $fitness_1 = 1/Z_1$, which is the reciprocal of the objective function in Step 1.

**Step 3**: Determine the number of iterations: If gen = *N*, output the optimal solution and go to Step 5; otherwise, go to Step 4.

**Step 4**: Perform selection, crossover, and mutation operations in order to obtain the offspring population. Set gen = gen + 1 and go to **Step 2**.

Selection: The selection process in this algorithm uses the roulette method and performs the crossover and mutation operation according to the crossover and mutation probability.

Crossover: First, randomly select two chromosomes from *M* chromosomes, then randomly select a gene position within the length of the gene sequence, and directly exchange the genes for the same position, as shown in Fig. 3. To complete the crossover operation, it is necessary to ensure that the second part satisfies constraints (6)-(9), and the third part satisfies constraints (8)-(10). Repeat the above process *M*/2 times until all individuals in the population are traversed.

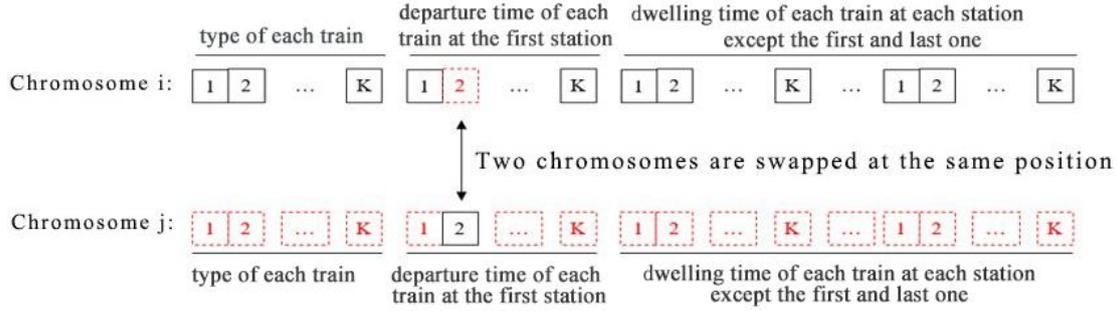

Fig 3. Chromosome Cross

Mutation: First select an arbitrary chromosome and randomly select a gene position to mutate. If the gene position is within the range of the first part, randomly select a gene from the $V$ integers, representing the train type, to replace the current position; if the gene position is in the second or third range, the number of genes at that position is randomly increased or decreased by a suitable value, as shown in Fig. 4. To complete the mutation operation, it is necessary to ensure that the second part satisfies constraints (6)-(9), and the third part satisfies constraints (8)-(10); otherwise, it must be mutated again until the constraints are satisfied. Repeat $M$ times until all individuals in the population are traversed.

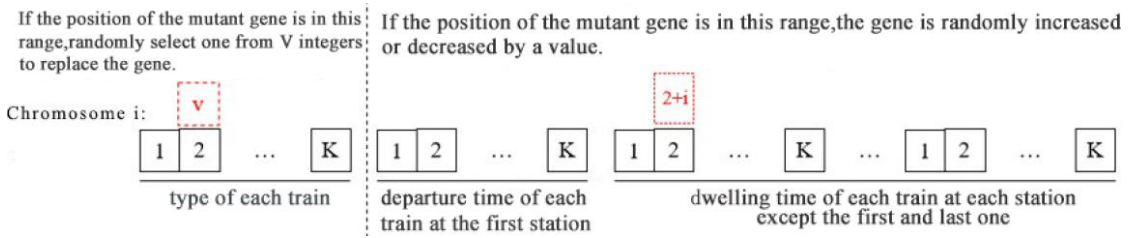

Fig 4. Chromosome Variation

**Step 5**: Using the optimal solution obtained in **Step 3** to execute the second GA.

Determine the population size $M$ and the maximal evolution algebra $N$ suitable for the second genetic algorithm, and initialize the iteration count gen=0. using real numbers to encode the chromosome, the gene sequence is shown in Fig. 5, and the randomly generated $K*(N-2)$ real numbers satisfying Eq. (21) are treated as chromosomes.

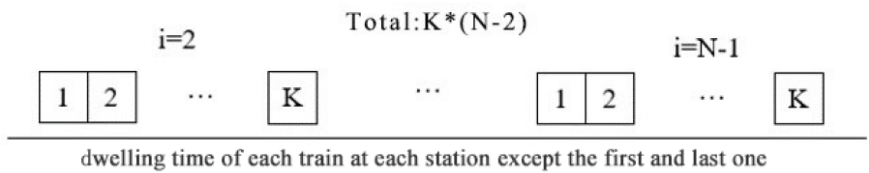

Fig 5. Chromosome coding in step 2

**Step 6**: Calculate the objective value and set the fitness function as $fitness_2 = 1/Z_2$.

**Step 7**: Determine the number of iterations: if gen = $N$, output the optimal solution and stop; otherwise, perform selection, crossover, and mutation operations to obtain the offspring population. Set gen = gen + 1 and go to Step 6.

The flowchart of the genetic algorithm is shown in Fig. 6.

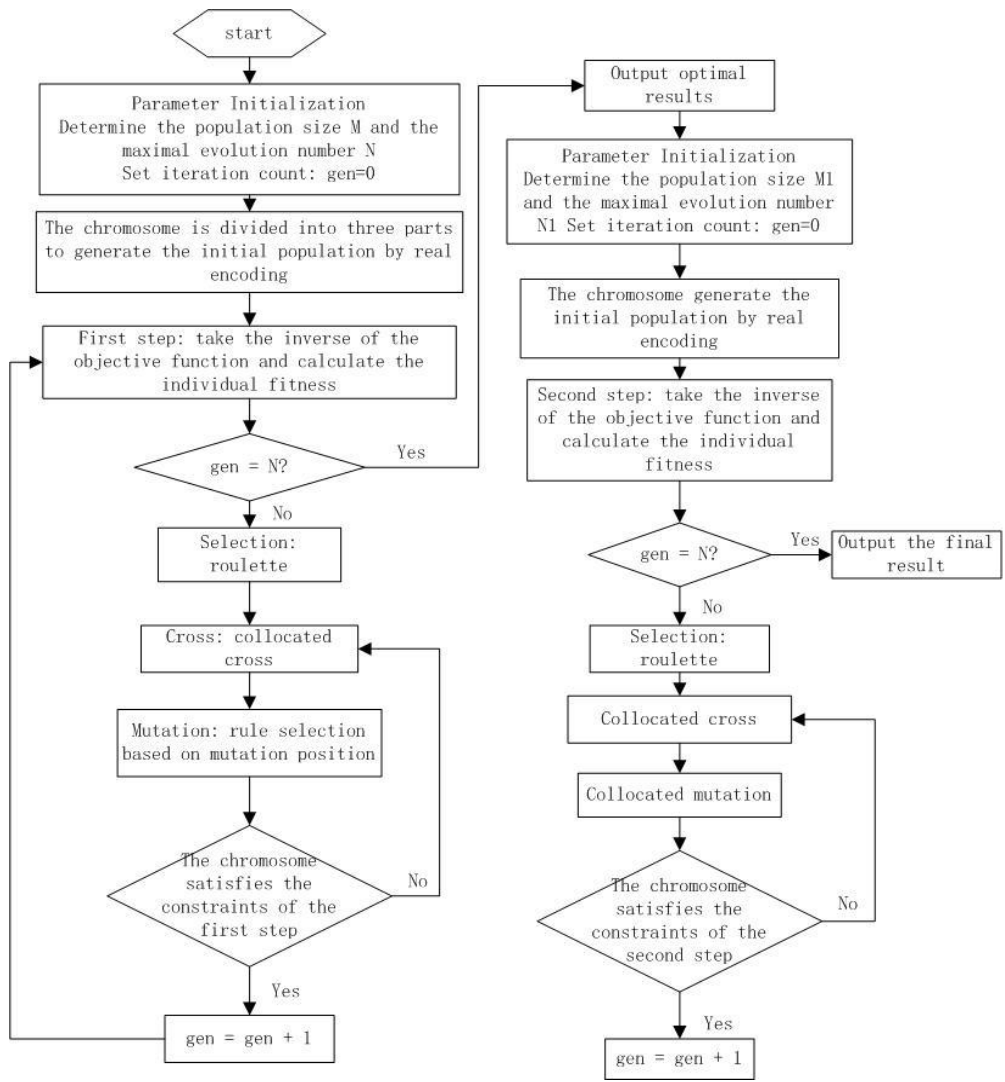

Fig 6. Flow chart of the genetic algorithm

## 4. Case analysis

Beijing Metro Line 9 connecting comprehensive transportation hub of Beijing West Railway Station is a vertical trunk line in the west of Beijing with total length of 16.5 kilometers. It runs north-south and has 13 stations, including 7 interchange stations. Each train of this line has 6 B-type carriages uniformly and a marshalling capacity of 1,440 pass. The maximal speed of the train is designed to 80 km/h. Trains are operated in different frequencies during a day. In morning and evening peak hours on weekdays, the headway is 4 minutes, while 6 minutes in off-peak hours. The dwelling time of train at each stop is 30s-45s. Passenger restriction measure is adopted at Beijing West Railway Station throughout the day, that is, the number of passengers entering the platform is controlled according to the congestion state.

This paper hopes to deal with fluctuations of coming passenger demand by combined operation of multi-marshalling and station control, so passenger restriction measures at the station is removed, which produces much waiting time outside the rail station. Assume that 60% of the passengers waiting for the subway at Beijing West Railway Station come from outer transportation, which is calculated from the data of coming outer transportation at the hub.

During [7:30, 9:10] of the morning peak hour, combining the passenger data from AFC every 5 minutes and the arrival of outer transportation at the hub, we obtain the passenger demand distribution of the Beijing Metro No. 9 at the hub as shown in Fig. 7.

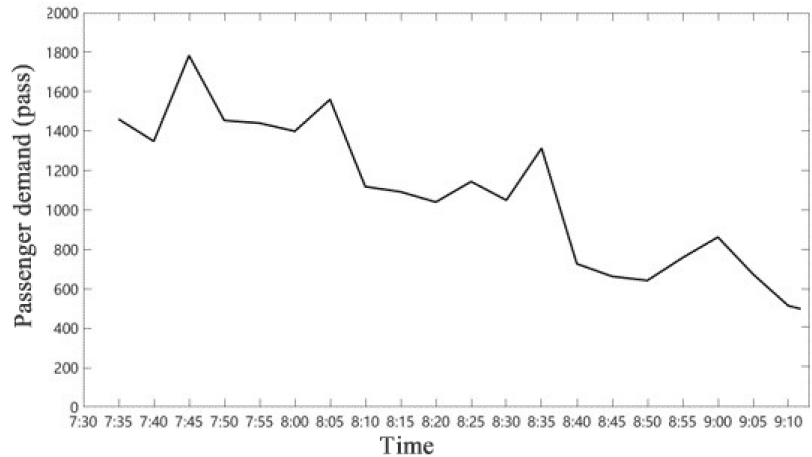

Fig 7. Passenger demand of Beijing Metro No. 9 at the hub

It can be seen from the figure that without current restrict measure, the demands fluctuate significantly for every 5 minutes and the overall trend is downward.

**4.1 Parameter setting**

The example line considers two types of train marshalling: 4 marshalling of 960 passengers and 8 marshalling of 1920 passengers. Consider the direction from National Library to Guogongzhuang Station and the headway is between 4.85 and 5.15 min. The running time data between stations are given in Table 1. The dwelling times of different trains at each station are optimized between 30 and 45 s.

The study period is [7:30, 9:10] which is divided into 20 time intervals. Based on the departure interval of rail transit trains and the total running time on the line as given in Table 1, we have 18 trains during this study period.

Table 1. Running time between stops

| Stations | National Library - Baishiqiao South | Baishiqiao South - Baiduizi | Baiduizi - Military Museum | Military Museum - Beijingxi Railway |
|---|---|---|---|---|
| Running time (min) | 1.6 | 1.4 | 2.7 | 2.1 |
| Stations | Beijingxi Railway - Liuliqiao East | Liuliqiao East - Liuliqiao | Liuliqiao - Qilizhuang | Qilizhuang - Fengtai East Street |
| Running time (min) | 1.7 | 1.9 | 2.6 | 1.9 |

| Stations | Fengtai East Street - Fengtai South Road | Fengtai South Road - Keyi Road | Keyi Road - Fengtai Science Park | Fengtai Science Park -Guogongzhuang |
|---|---|---|---|---|
| Running time (min) | 2.3 | 1.45 | 1.18 | 2.0 |

The second step deals with the random fluctuation of demand caused by the delay of outer transportation, which is represented by a random function. The outer transportation arriving at Beijing West Railway Station during the study period are numbered according to their scheduled arrival time, and then a random function $\xi(x,t)$ is used to indicate the delay situation, where $x$ is the No. of the delayed outer transportation, which is randomly selected; $t$ represents the delayed time randomly selected in (0, 30].

**4.2 Other operating strategies**

Denote the two-stage train operation strategy in this paper as full strategy (FS) and to give comparisons three train operation strategies are defined: strategy 1 (S1), strategy 2 (S2) and strategy 3 (S3) as shown in Table 2. These three operation strategies can be achieved by adjusting the optimization model in this paper.

Table 2. Operation strategies for comparisons

| Strategy | Marshalling | Station Control | Scheduled Headway |
|---|---|---|---|
| S1 | fixed | no | peak and off-peak |
| S2 | fixed | yes | peak and off-peak |
| S3 | variable | no | uniform |
| FS | variable | yes | uniform |

S1: The train marshalling is fixed to 6 and the first stage is omitted. The scheduled headway of 1-9 trains is fixed to 4 minutes, and that of 10-18 trains is fixed to 6 minutes. S1 is the current train operation strategy in use when the arrival of the outer transportation at the hub is fluctuated.

S2: The train marshalling is fixed of 6 and the first stage is omitted. The scheduled headway of 1-9 trains is fixed to 4 minutes, and that of 10-18 trains is fixed 6 minutes. The second state is the same as that of FS.

S3: It has only the first stage of marshalling optimization process, and the second stage is omitted.

In addition, since the second stage of the above 4 strategies is based on the consideration of the delay in the arrival of the Beijing West Railway Station, in order to compare the situation when no delay occurs, the S1 and FS are applied to the situation with no delay and are denoted as S1-N and FS-N, respectively.

Other parameters in GA includes: $M = 50$, $N = 500$, $P_c = 0.8$, and $P_m = 0.5$. Solve the above 6 cases and analyze the numerical results in the following section.

**4.3 Numerical results**

During the study period [7:30, 9:10], regardless the unstable operation results of the first 20 minutes, the calculation results of the six cases during time period [7:50, 9:10] are mainly focus on the matching of train supply and passenger demand, passenger waiting time, train travel time, and number of people left behind at the station.

(1) Matching of train supply and passenger demand

First, the matching of train supply with passenger demand is measured by the difference between the passenger demand and the number of passengers getting on the rail transit line in each unit period. The smaller the difference, the higher the degree of matching. In the period of [7:50, 9:10], passenger demand and the number of passengers boarding on the line at the hub under different strategies are both counted every 5 minutes and are drawn in Fig. 8 to show the difference.

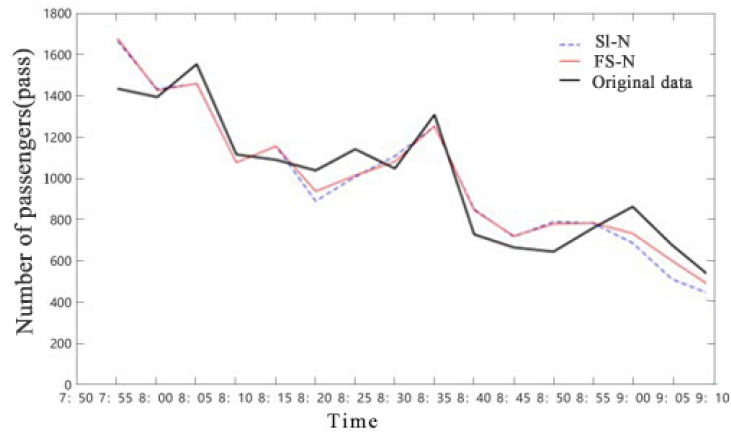

(a) S1-N and FS-N

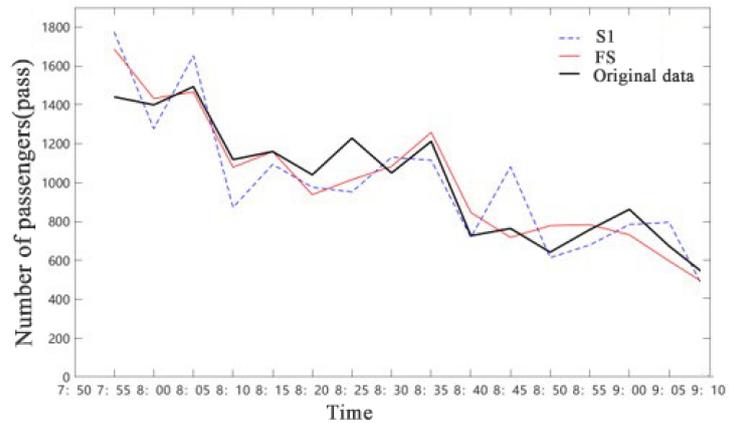

(b) S1 and FS

Fig 8. Passenger demand and train supply

Fig. 8 (a) is the results of S1-N and FS-N while Fig. 8 (b) is the results of S1 and FS. Fig. 8 (a) shows that if the outer transportation is operated strictly according to the schedule and has no fluctuation, peak and off-peak strategy (S1-N) can work almost as well as FS-N to supply the passenger demand at the hub, even though FS-N is a little better. But if the arrival time of outer transportation is not on schedule, FS shows much higher ability to deal with the fluctuation of passenger demand as shown in Fig. 8 (b). In fact, the delay of outer transportation exists indeed. If the delay information is released a relative long time beforehand, the train marshalling can be optimized to deal with this situation; if the delay information is obtained in real time, the station control of the train is much effective.,

(2) Passengers left behind

In this example, the numbers of passengers left behind after each train leaves each station are compared between four cases: S1-N, FS-N, S1 and FS, and the results are shown in Fig. 9. By comparing the figures from (a) to (d), we find that both with and without the fluctuated delay of outer transportation, FSs (FS-N and FS) do much better than S1s (S1-N and S1), which left much less passengers behind after the trains leave the stations. Moreover, high number of left passengers in all cases happens at later time period (corresponding to bigger train No.). The reason of S1s is the long headway of off-peak period while the reason of FSs is that the optimized train marshalling at later time period is small. On the other hand, high number of left passengers in all cases happens at Beijing West Railway Station and the nearby stations because of the high passenger demand at Beijing West Railway Station.

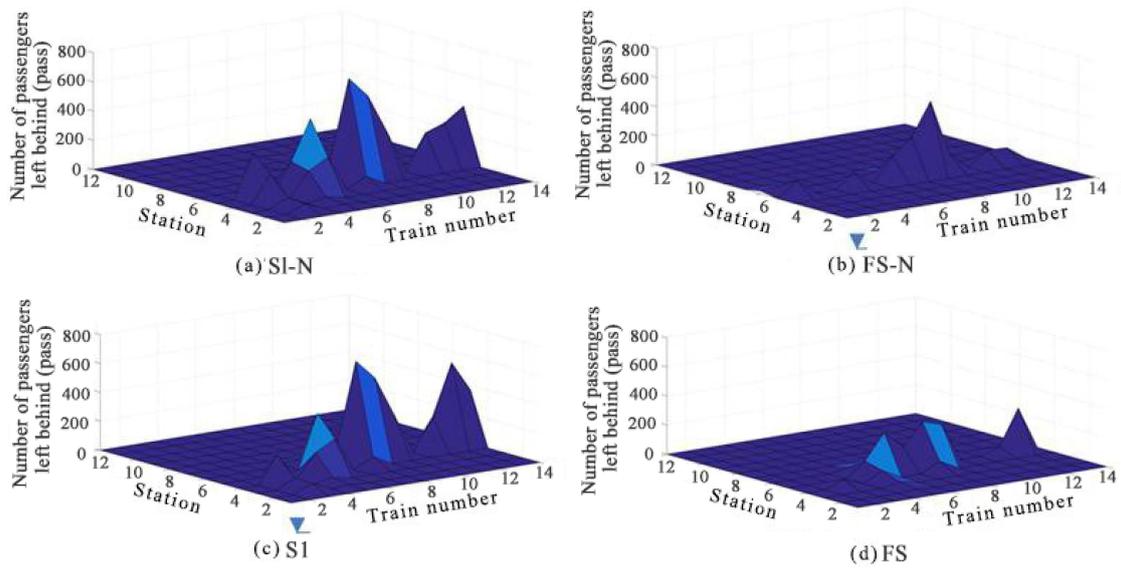

Figure 9. Number of passengers left behind

(3) Average passenger waiting time

During the period of [7:50, 9:10], the average waiting time per passenger are calculated every 5 minutes for the six cases and the results are shown in Table 3. The third column in the table represents the degree of improvement relative to the peak-and-off-peak departure interval strategy (corresponding S2-N or S2), and the fourth column represents the standard deviation of the average waiting time.

Table 3 Average waiting time

| Strategy | Average waiting time (min) | Improvement to S1-N/S1 | Standard deviation |
|---|---|---|---|
| S1-N | 3.092 | — | 0.93 |
| FS-N | 2.923 | 5.47% | 0.80 |
| S1 | 3.196 | — | 1.08 |
| S2 | 3.112 | 2.63% | 0.96 |
| S3 | 3.022 | 5.44% | 0.85 |
| FS | 2.935 | 8.17% | 0.84 |

According to the results in Table 3, all strategies have improvement in average waiting time per passenger compared to strategy of only peak-and-off-peak departure interval with and without outer transportation delay. If the outer transportation can arrive strictly according to the schedule, and the passenger demand is fixed, the average waiting time of FS-N has a reduction of 5.47% and a better stability (smaller standard deviation) compared to that of S1-N. If the outer transportation cannot arrive on time and have more or less delay, FS has even more advantage to reduce the average waiting time (a reduction of 8.17%) and to level off the standard deviation (from 1.08 to 0.84). S2 and S3 are better than S1 but worser than FS from these two respects. In a word, FS can do better both in reducing the total waiting time (same as average waiting time) and in averaging the waiting time to each passenger.

(3) Total travel time of the train

The total travel times of the 18 trains' whole trips in 4 strategies with and without outer transportation delay are shown in Fig. 10 to illustrate the effect of station control.

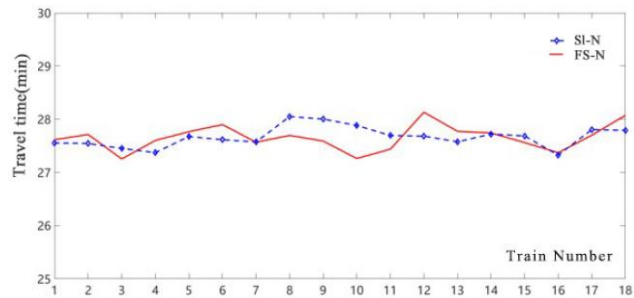

(a) S1-N and FS-N

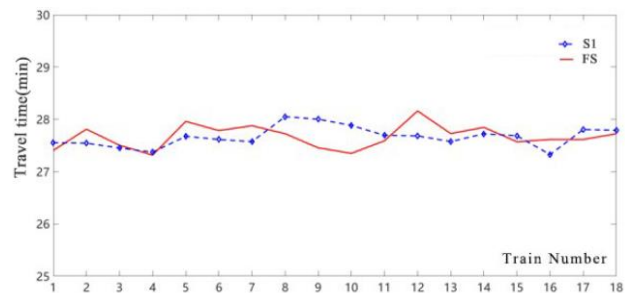

(b) S1 and FS

Figure 10. Travel time of train's whole trip

Comparing the results in both figures, because the station control of all the trains is from the point of passengers, some trains have smaller total travel time compared to scheduled travel time, while other trains have bigger total travel time. But all the total travel times concentrate at about 27.5 minutes, and the fluctuation among all trains is not apparent. The average travel time and standard deviation of each train for the 4 cases are shown in Table 4. From Table 4 we also find that if the outer transportation arrives on time, the strategy FS can even reduce the average travel time of the trains; on the other hand, if the outer transportation does not arrive on time, the average travel time of the trains has only increased 0.025 min, which means that FS remains a little change of the trains' travel time (also the on-board time of passengers) and obtain much reduced of the passengers' waiting time.

Table 4. Average travel time and standard deviation

| Strategy | Average travel time (min) | Standard deviation |
|----------|---------------------------|--------------------|
| S1-N     | 27.667                    | 0.189              |
| FS-N     | 27.653                    | 0.232              |
| S1       | 27.644                    | 0.232              |
| FS       | 27.669                    | 0.215              |

## 5. Conclusions

This paper analyzes the characteristics of passenger demand of urban rail transit lines connecting to integrated transportation hubs, which includes both the commute passengers along the lines and the passengers from the outer transportation. The passenger demand from the outer transportation is much relative to the arrival schedule of outer transportation; moreover, the outer transportation does not arrive on time due to kinds of reasons. Based on these characteristics of passenger demand, this paper establishes a combined two-stage model of train formation optimization and real-time station control. The goals are minimizing the total spare space on the train and minimizing the total passenger waiting time, where the first one is to make good use of the train resources and balance the congestion state of each train and the second one is to reduce passengers' travel costs. The two-stage model is solved by two GAs successively.

The main conclusions of this paper include:

The existed operation strategy of peak-and-off-peak headways only resolve the main fluctuation of the scheduled passenger demand, while the two-stage model in this paper resolve both the scheduled passenger demand and its usually delay from respects of both the train operation and the passengers. The first stage optimizes the multiple train formations at different periods and the second stage optimize the trains' dwelling time control at each station, where dwelling time can both shorter and longer than scheduled one.

This paper designs a solution method for the combined model of train marshalling and real-time station control based on the GA, which is realized by calling the GA twice. In the first stage, the GA calculates the number of muti-marshalling trains and the timetable, which is the input of the second stage; in the second stage, the GA calculates the optimized station control of the trains.

This paper considers 4 train operation strategies and 2 situations (depends on whether the outer transportation arrives on time or not at the hub) and does not consider the current passenger flow limiting measures at the station. The case analysis is based on Beijing Metro Line 9 connecting to

Beijing West Railway Station. Numerical results show that no matter whether the outer transportation arrives on time or not, after the train marshalling optimization and real-time station control, the train supply capacity and the passenger demand are more matched, the number of waiting passengers per train are reduced, and the waiting times of passengers arriving at different periods are more balanced. Moreover, the model can reduce passenger waiting time without increasing passengers' on-board time; at the same time, the current passenger flow limiting measures are unnecessary which increase the waiting time of passengers and have a large number of passengers strand in the hub station.

On the basis of the research in this paper, we can consider further discussions in the following aspects.

(1) When analyzing the impact of delays in the arrival of outer transportation at the hub station, the model considers only small delays. Therefore, the second stage of the model only adjusts the station control of the trains. However, when the delay is too long, and the passenger demand changes excessively, it should be reflected in the first stage of train marshalling optimization.

(2) It is necessary to consider real-time updating of all kinds of information that affect the passenger demand and utilized them in station control of trains in real-time and even train marshalling optimization.

## Acknowledgements

This work is jointly supported by the National Natural Science Foundation of China (71771019, 71621001, 71871130).